\documentclass[aps, prb, 12pt, letterpaper, reprint, superscriptaddress]{revtex4-2}

\usepackage[english]{babel}
\usepackage[utf8]{inputenc}
\usepackage{hyperref}
\usepackage{amsmath}
\usepackage{amssymb}
\usepackage{graphicx}
\usepackage{color}
\usepackage{bm}

\DeclareMathOperator{\Tr}{\mathrm{Tr}}
\DeclareMathOperator{\tr}{\mathrm{tr}}
\renewcommand{\Re}{\mathop{\mathrm{Re}}\nolimits}

\newcommand{\Sb}{\mathop{\mathrm{Tr}_{r}}\nolimits}
\newcommand{\Pj}{\mathop{\mathcal{P}_{t}}\nolimits}
\newcommand{\Qj}{\mathop{\mathcal{Q}_{t}}\nolimits}

\begin{document}

\title{Mode attraction in Floquet systems with memory: application to magnonics}

\author{Igor \surname{Proskurin}}
%\email{Igor.Proskurin@umanitoba.ca}
\affiliation{Department of Physics and Astronomy, University of Manitoba, Winnipeg, MB R3T~2N2, Canada}
\affiliation{Institute of Natural Sciences and Mathematics, Ural Federal University, Ekaterinburg 620002, Russia}

\author{Jephthah O. \surname{Iyaro}}
%\email{iyaroo@myumanitoba.ca}
\affiliation{Department of Physics and Astronomy, University of Manitoba, Winnipeg, MB R3T~2N2, Canada}

\author{Robert L. \surname{Stamps}}
\email{Robert.Stamps@umanitoba.ca}
\affiliation{Department of Physics and Astronomy, University of Manitoba,
Winnipeg, MB R3T~2N2, Canada}

\date{\today}

\begin{abstract}
Level attraction is a type of mode hybridization in open systems where instead of forming a hybridization gap, the energy spectrum of two modes coalesce in a region bounded by exceptional points. We demonstrate that this phenomenon can be realized in a Floquet system with memory, which appears in describing linear excitations in a nonlinear driven system with a limit cycle.  
Linear response of the system in this state is different from its response near thermodynamic equilibrium. We develop a general formalism and provide an example in the context of cavity magnonics, where we show that magnetic excitations in systems driven far from the equilibrium may show level attraction with cavity photons.  Our approach works equally well for quantum and semiclassical magnetic dynamics.  The theory is formulated so that it can be used in combination with micromagnetic simulations to explore a wide range of experimentally interesting systems.
\end{abstract}

\maketitle

\section{Introduction}
Excitations around dynamic steady states in open nonlinear systems  away from the equilibrium can show features that cannot be observed near the ground state, and can be broadly understood in terms of non-Hermitian physics \cite{Cao2015, Ashida2020}.  An example is level attraction as recently demonstrated in a dissipative magnon-polariton microwave cavity \cite{Harder2018}. This is a dynamic regime characterized by a region where the energy levels of the interacting cavity system coalesce \cite{Bernier2018}.  The appearance of exceptional points delineating the attraction is a non-Hermitian phenomenon \cite{Heiss2004, Heiss2012} that can take place in driven systems for some types of dissipation \cite{ElGanainy2018, Wang2020, Harder2021}.  The  energy levels near the exceptional points are sensitive to manipulation through external parameters and are potentially useful for mode control and sensing \cite{Chen2017, Hodaei2017, Zhong2019}.

Mode attraction and exceptional points have been studied extensively in magnonics both theoretically and experimentally \cite{Grigoryan2019, Rao2019, Wang2019, Yu2019_1, Boventer2019, Boventer2020, Tserkovnyak2020, Yuan2020,  Yang2020, Grigoryan2020, Rao2021,  Lu2021}.  Theoretical description has been based largely on various models of coupled oscillators borrowed from cavity electrodynamics \cite{Grigoryan2018, Proskurin2018, Proskurin2019, Peng2020} with additional non-Hermitian mechanisms \cite{Boventer2019, Boventer2020} such as dissipative coupling \cite{Xu2019, Yu2019} and nonlocal interactions \cite{Rao2020,Yao2019}.

Recently Floquet states for linear excitations around equilibrium in an open cavity magnonic system have been realized  experimentally \cite{Xu2020}. It has been demonstrated that higher order Floquet bands may contribute significantly to the cavity reflection spectrum in the Floquet ultrastrong coupling regime.

The main focus of the present paper is on magnonics around non-linear steady states away from equilibrium.  This allows us to realize level attraction, which does not appear in a linear Floquet cavity system. We show that dynamics of the system can be understood on the basis of a generalized Floquet theorem for non-Markovian kinetic equations \cite{Traversa2013, Magazzu2017, Magazzu2018, Traversa2020}. Regions of stability and instability, determined by the Floquet index of the system, can be associated with mode repulsion and attraction between the system and the excited states of the reservoir.  This approach is equally applicable to quantum and semi-classical dynamics, and can be used in combination with numeric methods.  Our theory is quite general, and can be applied to a variety of magnetic and non-magnetic systems.

In order to illustrate application of our theory, we demonstrate how the Floquet formalism can be used in magnonics by considering a microwave cavity loaded with a driven magnetic specimen.  This situation has been recently realized experimentally in a microwave cavity, where a magnetic specimen has been probed and driven out of equilibrium using separate ports \cite{Boventer2019, Boventer2020}.   In this system, the cavity photons serve as probes that can read out magnetization dynamics. Details of this process can be described using a generalized susceptibility \cite{Ono2019} given by a nonequilibrium Green function \cite{Zubarev1996, Tsuji2008, Aoki2014}.

\begin{figure}[t]
	\includegraphics[width=0.45\textwidth]{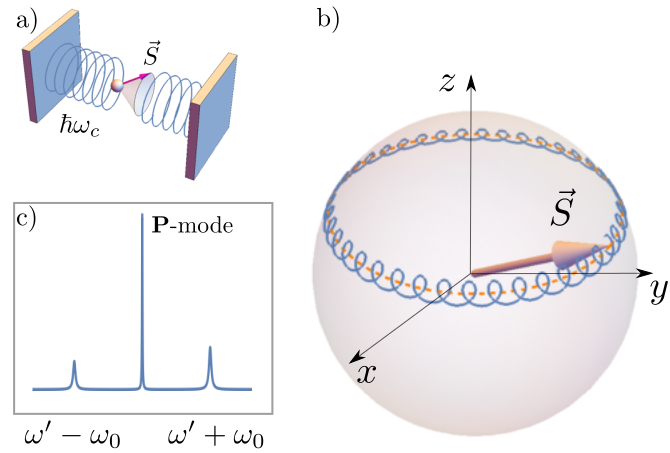}
	\caption{(a) A small magnetic specimen inside a microwave cavity resonator. (b) Schematic picture of an excitation around a \textbf{P}-mode stationary trajectory of spin  precession (dashed line), and (c) the corresponding Fourier spectrum showing the \textbf{P}-mode at the frequency of the driving field $\omega'$, and two side bands with the frequencies $\omega' \pm \omega_{0}$.}
\end{figure}

Far from thermodynamic equilibrium, nonlinear magnetization dynamics in systems that have uniaxial rotation symmetry can be characterized by steady state trajectories known as \textbf{P}-modes \cite{Bertotti2001}.  The number of \textbf{P}-modes is described by  the Poincar\'{e}-Bendixson theorem \cite{Hirsch2012} and their stability depends on details of microscopic interactions.  Stable \textbf{P}-modes provide a steady state about which elementary excitations can exists. These excitations have finite lifetime,  and form two side bands with respect to the driving frequency.  This feature has been used recently for nutation spectroscopy in nonlinear ferromagnetic resonance \cite{Li2019}.

We show that in  \textbf{P}-mode cavity magnonics a hybridization between cavity photons and excitations around the \textbf{P}-mode can occur, which depends on whether the cavity resonance is tuned to the lower or upper side band.  If the phase of the resonance in the lower band is shifted by $\pi$ with respect to the upper band level attraction instead of repulsion appears.

Our paper is organized as follows. In Sec.~\ref{SecII} we consider a general formalism for a probe system in contact with a driven reservoir. This formalism is applied to a cavity magnonic systems in Sec.~\ref{SecIII}. Section~\ref{SecIV} is reserved for the discussion of results, and Sec.~\ref{SecV} is for the conclusions.

\section{General formalism}
\label{SecII}
We begin by outlining a theory for periodically driven systems.  In the magnonic experiment described above, the cavity photon system is the probe, and is characterized by a density matrix $\hat{\rho}(t)$. The probe is assumed to be weakly coupled to the driven system (i.~e. the magnetic sample loading the cavity), which is characterized by the density matrix $\hat{\rho}_{r}(t)$.  The total Hamiltonian of the interacting system is given by
\begin{equation}
\hat{H}_{\mathrm{tot}}^{t} = \hat{H}_{s}^{t} + \hat{H}_{r}^{t} + \hat{H}_{i}^{t},
\end{equation}
where the first term is the Hamiltonian of the probe, the second corresponds to the driven system and the last is the interaction part. The index $t$ indicates that each term can be explicitly time-dependent. 

The total interacting system is characterized by the density matrix $\hat{\rho}_{\mathrm{tot}}(t)$, and its dynamics is described by the Liouville equation
\begin{equation} \label{liouv}
	\left( \frac{\partial}{\partial t} + i\mathcal{L}_{s}^{t} + i\mathcal{L}_{r}^{t} + i\mathcal{L}^{t}_{i} \right) \hat{\rho}_{\mathrm{tot}}(t) = -\varepsilon \left( \hat{\rho}_{\mathrm{tot}}(t) - \hat{\rho}(t) \hat{\rho}_{r}(t) \right), 
\end{equation}
where $i\mathcal{L}_{\alpha}^{t}\ldots = \frac{1}{i\hbar} [\ldots, \hat{H}_{\alpha}^{t}]$ ($\alpha = s,r,i$) denote the Liouville operators for the probe, driven system and the interaction part correspondingly.   We assume that  the probe and the driven system are uncoupled at $t \to -\infty$, so that the term on the right hand side is the boundary condition with $\varepsilon \to 0^{+}$, which breaks time reversal symmetry \cite{Zubarev1996}.

The density matrix of the probe system, which in this approach is treated as a reservoir, is obtained from  $\hat{\rho}_{\mathrm{tot}}(t)$ by taking a partial trace over the states of the driven system, $\hat{\rho}(t) = \Sb \hat{\rho}_{\mathrm{tot}}(t)$. The time-evolution of $\hat{\rho}_{r}(t)$ is teated as independent from dynamics of the probe system, and is described by the separate Liouville equation
\begin{equation} \label{lvl}
	\frac{\partial \hat{\rho}_{r}(t)}{\partial t} + i\mathcal{L}_{r}^{t}\hat{\rho}_{r}(t) = 0.
\end{equation}

A closed master equation for $\hat{\rho}(t)$ can be derived using the methods of relaxation dynamics in open dissipative systems  \cite{Zubarev1996}. We outline the most important steps in Appendix~\ref{app_A}. For weakly coupled systems, the master equation  for $\hat{\rho}(t)$ has the following form
\begin{multline} \label{mas}
\frac{\partial \hat{\rho}(t)}{\partial t} 
+ \frac{1}{i\hbar} [\hat{\rho}(t), \hat{H}_{s}^{t} ] 
= \frac{1}{(i\hbar)^{2}} \int_{-\infty}^{t} dt' e^{-\varepsilon(t - t')} \\
\times \Tr_{r}[\hat{H}_{i}^{t}, [\hat{H}_{i}^{t'}(t,t'),\hat{\rho}(t,t')\hat{\rho}_{r}(t)]],
\end{multline}
where $\hat{H}_{i}^{t'}(t,t') = \hat{U}_{tt'}\hat{H}_{i}^{t'}\hat{U}_{tt'}^{-1}$ and $\hat{\rho}(t,t') = \hat{U}_{tt'}\hat{\rho}(t')\hat{U}_{tt'}^{-1}$ are defined with the evolution operator
\begin{equation} \label{Uev}
\hat{U}_{tt'} = \mathcal{T} \exp\left[ -\frac{i}{\hbar} \int_{t'}^{t} d\tau
\left( \hat{H}_{s}^{\tau} + \hat{H}_{r}^{\tau} \right) \right].
\end{equation}
Here $\mathcal{T}$ denotes the time-ordering operator and the trace in the right hand side of Eq.~(\ref{mas}) is with respect to the Hilbert space of the driven system.

In the Markov's approximation, $\hat{U}_{tt'} \hat{\rho}(t') \hat{U}^{-1}_{tt'} \approx \hat{\rho}(t)$, and Eq.~(\ref{mas}) reduces to the Lindblad master equation \cite{Zhao2021}.  It is important to keep track of memory effects, which later allow us to calculate the energy spectrum of the interacting system.

\subsection{Kinetic equation for the probe system}
The right hand side of Eq.~(\ref{mas}) is simplified if the interaction part is taken as a product of operators, $\hat{H}_{i} = \hat{a}\hat{F} + \hat{a}^{\dag}\hat{F}^{\dag}$, where $\hat{a}$ and $\hat{a}^{\dag}$ characterize the probe system, and $\hat{F}$ and $\hat{F}^{\dag}$ act entirely in the space of the driven system.  In the case when $\hat{a}$ and $\hat{a}^{\dag}$ satisfy a boson commutation relation, Eq.~(\ref{mas}) yields the following non-Markovian kinetic equation for the field amplitude of the probe system $\langle \hat{a} \rangle^{t} = \Tr(\hat{a}\hat{\rho}(t))$
\begin{equation} \label{kin}
i\hbar\frac{d \langle \hat{a} \rangle^{t}}{dt} +
\langle [\hat{H}_{s}^{t}, \hat{a}] \rangle^{t} 
- \int_{-\infty}^{t} dt' G(t,t') 
\langle \hat{a} \rangle^{t'} = \langle \hat{F}^{\dag}(t) \rangle_{r},
\end{equation}
where the memory kernel is given by the nonequilibrium retarded Green function \cite{Zubarev1996, Kamenev2009, Aoki2014}
\begin{equation} \label{rgf}
G(t,t') = \frac{1}{i\hbar} e^{-\varepsilon(t -t')} \theta(t -t')
\langle [\hat{F}^{\dag}(t), \hat{F}(t')] \rangle_{r}.
\end{equation}
Here the operators $\hat{F}^{\dag}(t)$ and $\hat{F}(t)$ are in the Heisenberg picture and satisfy the equation of motion $i\hbar\partial_{t} \hat{F} = [\hat{F}, \hat{H}_{r}^{t}]$ with the time-dependent Hamiltonian. The average in Eq.~(\ref{rgf}), $\langle \ldots \rangle_{r} \equiv \Tr_{r}[\ldots \hat{\rho}_{r}(-\infty)]$, is taken with respect to the density matrix of the driven system at the initial time moment, $\hat{\rho}_{r}(-\infty)$.  The last term in Eq.~(\ref{rgf}) is the driving inherited from the dynamics of the reservoir.

\subsection{Semi-classical periodically driven systems}
For a semi-classical nonlinear system near the stable limit cycle regime of motion with period $T$, the memory kernel in Eq.~(\ref{rgf}) is bi-periodic in time, such that $G(t+T, t'+T) = G(t, t')$. To illustrate this, we expand the operators in the interaction Hamiltonian near the steady state trajectory using $\hat{F}(t) = F_{\mathrm{cl}}(t) + \delta \hat{F}(t)$ \cite{Kamenev2009}, where the first term describes the semi-classical solution of the equations of motion, and $\delta \hat{F}(t)$ is a perturbation.  Requiring that the elementary excitations around the steady state be characterized by one degree of freedom,  we expand $\delta \hat{F}(t)$ around $F_{\mathrm{cl}}(t)$ as
\begin{equation}
\delta \hat{F}(t) = \left(\frac{\partial F_{\mathrm{cl}}}{\partial b}\right)_{t} \hat{b}(t) + \left(\frac{\partial F_{\mathrm{cl}}}{\partial b}\right)_{t} \hat{b}^{\dag}(t),
\end{equation}
taken to the first order in terms of the boson operators $b(t)$ and $b^{\dag}(t)$, which describe the excitations.  In the linear approximation, $b(t)$ and $b^{\dag}(t)$ satisfy equations of motion with time-periodic coefficients. These can be characterized by a Floquet solution $(\hat{b}(t), \hat{b}^{\dag}(t))^{T} = \hat{A}e^{i\nu t} \bm{F}_{\nu}(t) + \hat{B}e^{-i\nu t} \bm{F}^{*}_{\nu}(t)$, where $\nu$ is the Floquet index and $\bm{F}_{\nu}(t)$ is periodic in time with period $T$. The coefficients $\hat{A}$ and $\hat{B}$ are determined from the initial conditions $\hat{b}(0) = \hat{b}$ and $\hat{b}^{\dag}(0) = \hat{b}^{\dag}$.

By substituting $b(t)$ and $b^{\dag}(t)$ into the memory kernel (\ref{rgf}), we find $G(t,t') \sim i[K_{1\nu}(t)K_{1\nu}^{*}(t')\exp(i\nu(t-t')) - K_{2\nu}(t)K_{2\nu}^{*}(t')\exp(-i\nu(t-t'))]$ ($t>t'$), where $K_{i\nu}(t)$ ($i=1,2$) is a periodic function whose explicit form depends on the details of the limit cycle and interactions. This form of the memory kernel is manifestly bi-periodic in time so that the kinetic equation (\ref{kin}) falls under the conditions of the generalized Floquet theorem \cite{Traversa2013} and can be analyzed by methods of embedding \cite{Magazzu2017} and harmonic balance \cite{Traversa2020}. This memory kernel can be also interpreted as a linear susceptibility around the nonlinear steady state in the Floquet system, similar to Ref.~\cite{Ono2019}.

By applying the generalized Floquet theorem \cite{Traversa2013, Magazzu2017} to the homogeneous equation associated with Eq.~(\ref{kin}), we find that dynamics the probe system can be characterized by a Floquet index $\lambda$, which depends on parameters of the driven system and interactions \cite{Traversa2020}. If we take the probe system in a form of Harmonic oscillator, $\hat{H}_{s} = \hbar\omega_{c}\hat{a}^{\dag}\hat{a}$ with the frequency $\omega_{c}$, the Floquet index $\lambda = \lambda(\omega_{c}, \nu)$ becomes a function of $\nu$ and $\omega_{c}$, and can be considered as the hybridized spectrum of the interacting system.

In the absence of driving, when $\nu$ is the frequency of the excitations around equilibrium, hybridization between two energy levels leads to a level repulsion, so that $\lambda(\omega_{c}, \nu)$ remains real outside the hybridization gap. For a driven Floquet system it is possible, however, that $\lambda$ may become complex even if the systems are characterized by real $\omega_{c}$ and $\nu$ in absence of interaction.

To illustrate this idea, let us consider the case when time-dependence of the coefficients $K_{i\nu}(t)$ in the memory kernel can be approximated by a single harmonic,
$K_{i\nu}(t) = K_{i\nu}e^{i\omega't}$, where $\omega'$ is associated with the driving frequency.  As we show later, this case is realized for a magnetic oscillator driven with the circularly polarized field. The Green function in this case becomes a function of  $t - t'$, and has the following form
\begin{multline} \label{sus}
	G(t-t') = \frac{e^{i\varepsilon(t-t')}}{i\hbar\nu}\theta(t-t') \\
	\times \left(|K_{1\nu}|^{2} e^{i(\omega' + \nu)(t-t')} - |K_{2\nu}|^{2} e^{i(\omega' - \nu)(t-t')}\right),
\end{multline}
which shows two resonances with the negative and positive frequencies with respect to reference frequency $\omega'$. Note that these resonances enter with opposite signs that represents an additional $\pi$ phase difference.

The energy spectrum of the coupled system is found by Fourier transforming Eq.~(\ref{kin}), which leads to  $\omega - \omega_{c} = G(\omega)$, where $G(\omega) = \int_{-\infty}^{\infty}d\tau \exp(-i\omega\tau) G(\tau)$.  The first term in Eq.~(\ref{sus}) describes level hybridization between $\omega_{c}$ and $\omega'+\nu$ with the hybridization gap proportional to $|K_{\nu}|$. The second term has $\omega' - \nu$ and corresponds to level attraction when $\omega'$ is greater that $\nu$.

\section{Application to magnonics}
\label{SecIII}
For magnonics, we associate $\hat{H}_{s}$ with a system of cavity photons, and $\omega_{c}$ with the resonant frequency of cavity. The photons interact with a spin system, which we model as a driven reservoir as above. Spin dynamics of the reservoir are described semi-classically. The interaction between the photon and spin system is assumed to be of the form of a dipolar interaction, $\hat{H}_{i} = g (\hat{a} \hat{S}^{(+)} + \hat{a}^{\dag} \hat{S}^{(-)})$ where $g$ is the interaction constant. The operators $\hat{S}^{(\pm)}=\hat{S}^{x} \pm \hat{S}^{y}$ denote the circular components of the spin and $\hat{a}$ ($\hat{a}^{\dag}$) is the photon annihilation (creation) operator. The weak coupling assumption in Eq.~(\ref{kin}) means that interaction with cavity photons does not affect the steady state magnetization dynamics. Consequently, $g$ should be small compared to the characteristic energy of the ferromagnetic resonance.

The kinetic equation in Eq.~(\ref{kin}) works equally well for quantum and classical dynamics of the reservoir. In the latter case, one has to replace the commutator with the Poisson bracket, $(i\hbar)^{-1}[\hat{F}^{\dag}(t), \hat{F}(t)] \to \lbrace F^{*}(t), F(t) \rbrace$ \cite{Zubarev1996}, where $F(t)$ and $F^{*}(t)$ are functions of canonical variables, and the trace over the Hilbert space becomes an integral over phase space.

For a block spin system, semi-classical dynamics is described by the Lagrangian $\mathcal{L} = \sum_{i} S \cos\theta_{i} \partial \phi_{i}/\partial t - \mathcal{H}_{m}$ where the first term is the Berry phase expressed in terms of the azimuthal angle, $\phi_{i}$, and polar angle, $\theta_{i}$, and the last term is the spin Hamiltonian. The Poisson bracket for the two block spin components is defined as  \cite{Fogedby1980}
\begin{multline} \label{psn}
	\left\lbrace S^{\alpha}(t), S^{\beta}(t') \right\rbrace = -\frac{1}{S\sin\theta} \\ \times
	\left(\frac{\partial S^{\alpha}(t)}{\partial \phi_{1}} \frac{\partial S^{\beta}(t')}{\partial \theta_{1}} - \frac{\partial S^{\alpha}(t)}{\partial \theta_{1}} \frac{\partial S^{\beta}(t')}{\partial \phi_{1}} \right),
\end{multline} 
where the derivatives are taken with respect to $\phi_{1} = \phi(t_{1})$ and $\theta_{1} = \theta(t_{1})$ at the initial time $t_{1}$ and are treated as the initial conditions for the spin trajectory $\bm{S}(t) = \bm{S}(t; \phi_{1}, \theta_{1})$.  For $t=t'$, this expression is evaluated as $\lbrace S^{\alpha}, S^{\beta} \rbrace = \epsilon_{\alpha\beta\gamma}S^{\gamma}$, which is a semi-classical analog of the spin commutation relations.

\subsection{Semi-classical spin driven by the circularly polarize magnetic field}
The spin dynamics is calculated from the Landau-Lifshitz-Gilbert equation
\begin{equation} \label{llg}
	\frac{\partial \bm{S}}{\partial t} = -\gamma \bm{S} \times \bm{B}_{\mathrm{eff}} + \frac{\alpha}{M_{s}} \bm{S} \times \frac{\partial \bm{S}}{\partial t},
\end{equation} 
where $\gamma$ is the gyromagnetic ratio, $\alpha$ is the Gilbert damping constant, $M_{s}$ is the saturation magnetization, and $\bm{B}_{\mathrm{eff}} = -\delta \mathcal{H}_{m}/\delta \bm{S}$ is the effective field.  Here, we consider uniform precession of a single block macro-spin driven with a circularly polarized magnetic field in a situation where the rotation symmetry along the $z$ axis is preserved.  This configuration supports existence of the time-harmonic \textbf{P}-modes and prevents the onset of a chaotic regime \cite{Bertotti2001}.

We transform the equation of motion in Eq.~(\ref{llg}) to dimensionless form by introducing the following notations
\begin{equation}
	\bm{m} = \frac{\bm{S}}{M_{s}}, \quad \bm{h}_{\mathrm{eff}} = \frac{\bm{B}_{\mathrm{eff}}}{\mu_{0}M_{s}}, \quad  
	\tilde{t} = \gamma \mu_{0} M_{s}t,
\end{equation}
where $\mu_{0}$ is the vacuum permeability, and $\tilde{t}$ is the dimensionless time variable. In the dimensionless units, the Eq.~(\ref{llg}) becomes $ 	\dot{\bm{m}} = - \bm{m} \times \bm{h}_{\mathrm{eff}} + \alpha \bm{m} \times \dot{\bm{m}} $, where we identify four contributions to the effective field, $ \bm{h}_{\mathrm{eff}} = \bm{h}_{a} + \bm{h}_{M} + \bm{h}_{\mathrm{AN}} + \bm{h}_{\mathrm{ex}} $. The first  is the applied field, which contains a static field along the $z$ axis and the transverse dynamic driving field, $\bm{h}_{a} = h_{az} \hat{\bm{z}} + \bm{h}_{\perp}(t)$.  The second term is the demagnetizing field that preserves the rotation symmetry along $z$, $ \bm{h}_{M} = -N_{\perp} \bm{m}_{\perp} - N_{z} m_{z} \hat{\bm{z}} $. The third term is a uniaxial anisotropy field along the $z$ direction, $ \bm{h}_{\mathrm{AN}} = 2K_{1} m_{z} \hat{\bm{z}}/(\mu_{0}M_{s}^{2}) $ with $K_{1}$ being the anisotropy constant. And finally, since we only consider dynamics of uniformly magnetized medium, the exchange field $\bm{h}_{ex}$ vanishes.  The total effective field is written as
$ \bm{h}_{\mathrm{eff}} = \bm{h}_{\perp}(t) + (h_{az} + \varkappa m_{z} )\hat{\bm{z}}$ where $\varkappa = 2K_{1}/(\mu_{0}M_{s}^{2}) + N_{\perp} - N_{z}$ \cite{Bertotti2001}. The physical fields are $\bm{B}_{\perp} = \mu_{0}M_{s} \bm{h}_{\perp} $ and $B_{z} = \mu_{0} M_{s} h_{az}$.

The equations of motion take the most simple form in a frame of reference co-rotating with the driving field \cite{Rabi1954}. We take the driving field as $\bm{h}_{\perp}(t) = h_{a\perp} (\cos \omega' t, \sin \omega' t, 0)$, and use the following parametrization for the magnetization,  $\bm{m} = [
\cos ( \omega' t - \phi)\sin \theta, \sin (\omega' t - \phi) \sin \theta, \cos \theta]$,  where $\phi = \phi(t)$ and $\theta = \theta(t)$ are dynamic variables. In this parametrization, from Eq.~(\ref{llg}) we obtain the following system of autonomous differential equations \cite{Hirsch2012} on the surface of a sphere \cite{Bertotti2001}
\begin{eqnarray} \label{eom10}
	\dot{\theta} + \alpha \sin \theta \dot{\phi} &=& \varkappa \left[b_{\perp} \sin \phi - \Omega \sin \theta\right], \\ \label{eom11}
	\alpha \dot{\theta} + \sin \theta \dot{\phi}  &=& \varkappa \left[b_{\perp} \cos\phi\cos\theta -\sin \theta \left(b_{z} + \cos \theta \right)\right],
\end{eqnarray}
where  $b_{\perp} = h_{a\perp}/\varkappa$, $b_{z} = (h_{az} - \tilde{\omega}')/\varkappa$, $\Omega = \alpha \tilde{\omega}'/\varkappa$, and $\tilde{\omega}' = \omega' / (\gamma \mu_{0} M_{s}) $ denotes the dimensionless frequency.

The static solution of these equations can be conveniently parameterized as
\cite{Bertotti2001}
\begin{eqnarray} \label{st10}
	b_{z} &=& m_{z} \left( v - 1\right), \\ \label{st11}
	b_{\perp}^{2} &=& \left(1 - m_{z}^{2}\right)\left(\Omega^{2} + v^{2}\right).
\end{eqnarray}
where $m_{z} = \cos\theta$, and $v = \Omega \cot\phi$. 
In the original frame, these solutions correspond to uniform magnetization precession with frequency $\omega'$, and are known as ``\textbf{P}-modes''. Stability of the \textbf{P}-modes has been studied in Ref.~\cite{Bertotti2001}.

In the case of $\alpha = 0$ and $\varkappa = 0$, these equations reduce to $	\sin \phi = 0$, and $\tan\theta  = b_{\perp}/b_{z}$,
which describe the magnetization aligned along the direction of the stationary effective field $b_{\perp}\hat{\bm{x}} + b_{z}\hat{\bm{z}}$ in the corotating frame \cite{Rabi1954}.

\begin{figure*}
	\centerline{\includegraphics[width=1.0\textwidth]{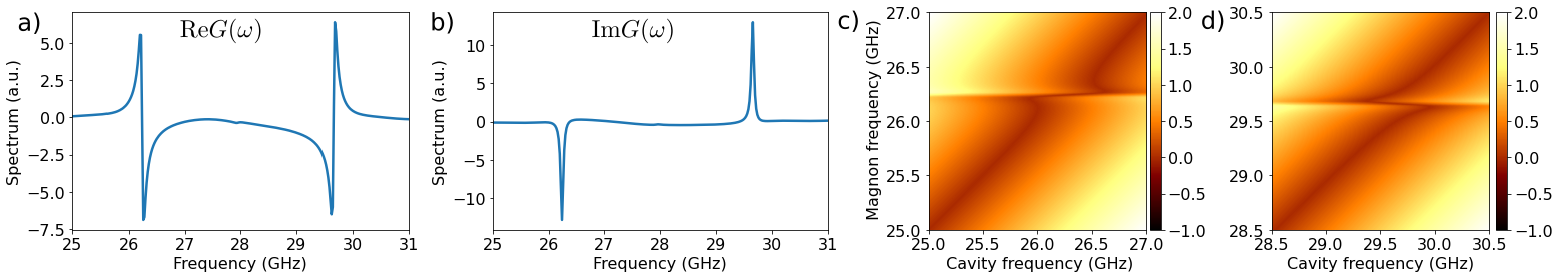}}
	\caption{Real (a) and imaginary (b) parts of $G(\omega)$ obtained in micromagnetic simulations for a spherical particle with $B_{z} = 1$~T, $B_{\perp} = 0.01$~T, and $\omega'/2\pi = 27.95$~GHz. Two bands correspond to the excitations with the frequencies $\omega' \pm \omega_{0}$. The density plots for level attraction near $\omega' - \omega_{0}$ (c) and repulsion near $\omega' + \omega_{0}$ (d) are obtained from the equation $\omega - \omega_{c} = \Re G(\omega)$ with $\tilde{g}/2\pi = 35$~MHz. }
	\label{fig2}
\end{figure*}

Linear excitations around a stable \textbf{P}-mode can described with expansions $\theta(t) = \theta_{0} + \delta \theta(t)$ and $\phi(t) = \phi_{0} + \delta \phi(t)$, which leads to the following equations of motion in the co-rotating frame 
\begin{equation} \label{eom}
\begin{pmatrix}
\delta \dot{\theta} \\
\delta \dot{\phi  }
\end{pmatrix}	 
	 = 
	\frac{\varkappa}{1 + \alpha^{2}}
	\begin{pmatrix}
		\alpha(1 - m_{z}^{2}) - A& B \sin\theta_{0} \\
		\dfrac{1 - m_{z}^{2} -B}{\sin\theta_{0}} & -A
	\end{pmatrix}
	\!
\begin{pmatrix}
\delta \theta \\
\delta \phi  
\end{pmatrix},
\end{equation}
where $A=\alpha v + \Omega m_{z}$ and $B = v - \alpha\Omega m_{z}$. From these equations of motion, we calculate the eigenfrequencies for linear excitations $\tilde{\lambda}_{0}= -\tilde{\gamma}_{0} \pm i\tilde{\omega}_{0}$, where $\tilde{\gamma}_{0}$ and $\tilde{\omega}_{0}$ are in the dimensionless units, e.~g., $\tilde{\omega}_{0}=\omega_{0}/(\gamma\mu_{0}M_{s})$. Here, $\omega_{0}$ corresponds the same frequency with the physical dimension restored. These frequencies can be found a general form, but in order to avoid complicated expressions, we only present results in the absence of uniaxial anisotropy,  i.e. $\varkappa \to 0$. This gives $\tilde{\gamma}_{0} = \alpha[b_{z}/m_{z} + \tilde{\omega}'m_{z}]/(1+\alpha^{2})$ and $\tilde{\omega}_{0} = [b_{z}/m_{z} - \alpha^{2}\tilde{\omega}'m_{z}]/(1+\alpha^{2})$.  A general analysis is qualitatively the same and can be found in Appendix~\ref{app_B}.

Next, we apply our general formalism to the spin-photon interactions inside the microwave cavity.  The Green function in Eq.~(\ref{rgf}) in this case is reduced to the spin-spin Poisson bracket $G(t,t') \sim \left\lbrace S^{(-)}(t), S^{(+)}(t') \right\rbrace$ ($t>t'$). In the linear approximation,  this corresponds to a linear susceptibility around the \textbf{P}-mode. The corresponding Poisson bracket is calculated from Eq.~(\ref{eom}) by solving the equations of motion and taking the derivatives with respect to the initial conditions.  This gives
\begin{multline} \label{ps2}
	\left\lbrace S^{(-)}(t), S^{(+)}(t') \right\rbrace =              
	2ig^{2}e^{-(\gamma_{0}+i\omega')(t-t')} \\
	\left(\cos^{4}\frac{\theta_{0}}{2}e^{-i\omega_{0}(t-t')} - \sin^{4}\frac{\theta_{0}}{2}e^{i\omega_{0}(t-t')}\right),
\end{multline}
in agreement with Eq.~(\ref{sus}). This expression shows two side bands around $\omega'$ with the frequencies $\omega'+\omega_{0}$ and $\omega'-\omega_{0}$, which correspond to linear excitations around the \textbf{P}-mode.  We note that  this situation has been experimentally observed in Ref.~\cite{Li2019}. When the spin is driven well out of equilibrium, the intensities of both side bands are the same, while close to thermodynamic equilibrium, $\theta_{0} \approx 0$,  the lower side band disappears as $\theta^{4}_{0}$, and the upper side band evolves into the usual ferromagnetic resonance with the frequency $\gamma B_{z}$.

\subsection{Level attraction with cavity photons outside of equilibrium}
Since the left hand side in Eq.~(\ref{ps2}) depends only on $t - t'$, the energy spectrum of the coupled spin-photon systems can be found by Fourier transforming Eq.~($\ref{kin}$) and solving the equation $\omega - \omega_{c} = \Re G(\omega)$, where $G(\omega) = \int_{-\infty}^{0} G(\tau) \exp(i\omega \tau) d\tau $.
When the frequency of the cavity mode is close to the frequency of the lower side band, $\omega'-\omega_{0}$, the energy spectrum determined from this equation is given by
\begin{equation}
	\omega^{(\pm)} = \frac{\omega_{c} + \Delta}{2} \pm \sqrt{\left(\frac{\omega_{c} - \Delta}{2}\right)^{2} - \tilde{g}^{2}},
\end{equation}
where $\Delta = \omega' - \omega_{0}$ is the frequency of the lower side band and $\tilde{g} = g\sin^{2}(\theta_{0}/2)$ is the effective coupling parameter for the level attraction. Level attraction occurs in the region $(\omega_{c} - \Delta)^{2} < 4\tilde{g}^{4}$. The coupling parameter is renormalized by the spin precession angle, and, therefore,  strongly depends on the amplitude of the driving field. Close to thermodynamic equilibrium, the $ \tilde{g}$ disappears as $\theta_{0}^{2} \sim [\gamma B_{\perp}/(\gamma B_{z} - \omega')]^{2}$. 

Strong enhancement of the effective coupling $\tilde{g}$ to the lower side-band with driving may be considered as an analog of Floquet ultra-strong coupling in Ref.~\cite{Xu2020}. The main difference, however, from the Floquet states near equilibrium is that higher order harmonics are not excited even for strong driving. This allows for control of the coupling parameter through the steady state avoiding contributions from higher harmonic modes.

\section{Numerical results and discussion}
\label{SecIV}
An advantage of considering semi-classical spin dynamics is that the nonequilibrium Green function for the spin components in Eq.~(\ref{kin}) can be computed numerically from Eq.~(\ref{psn}) for cases of practical interest. To illustrate, we performed micromagnetic simulations of macrospin dynamics using the \texttt{mumax3} package \cite{Vansteenkiste2014}. We considered dynamics of $16 \times 16 \times 16$ ellipsoid particle with the diameter of $100$~nm, $M_{s} = 1\times 10^{6}$~A/m, the exchange stiffness $A_{\mathrm{ex}} = 15\times10^{-12}$~J/m, and $\alpha = 9\times10^{-3}$ with enabled demagnetizing fields. The the static field $B_{z} = 1$~T has been applied along the $z$ axis, and the circularly polarized microwave field with the frequency $\omega'/2\pi = 27.95$~GHz and  $B_{\perp} = 0.01$~T has been applied perpendicular to $z$. The initial conditions for the \textbf{P}-mode have been identified from the stationary precession after a simulation time of $50$~ns. The Poisson bracket has been evaluated by identifying the \textbf{P}-mode precession and estimating the derivatives in Eq.~(\ref{psn}) numerically for the linear regime of deviation from the \textbf{P}-mode trajectory.

The results of simulations, presented in Fig.~\ref{fig2} (a) and (b), are in qualitative agreement with the analytical solution in Eq.~(\ref{ps2}) when demagnetizing fields are enabled in simulations. Without the demagnetizing field, the agreement becomes quantitative.

In the frequency domain, the Poisson bracket in Fig.~\ref{fig2}~(a) shows two side bands around the driving frequency $\omega'$. Note that the sign on the real part of the lower side band has been inverted with respect to the sign of the upper side band, which corresponds to $\pi$ phase shift between two lines, in agreement with Eq.~(\ref{ps2}).  We note that the coupling of the cavity photons to the excitations in the lower side band can be effectively described in terms of a non-Hermitian Hamiltonian, $\hat{\mathcal{H}} = \omega_{c}\hat{a}^{\dag}\hat{a} + \Delta \hat{b}^{\dag}\hat{b} + g_{\mathrm{eff}}e^{i\varphi} (a^{\dag}b + ab^{\dag} )$, with $ \varphi = \pi/2$, and $\hat{b}$ and $\hat{b}^{\dag}$ being magnon ladder operators. The possibility of such ``dissipative'' coupling has been discussed in a different context in Refs.~{\cite{Grigoryan2019, Boventer2020}. The density plots for attraction and repulsion, computed numerically  from the equation $\omega-\omega_{c} = G(\omega)$, are shown in Fig.~\ref{fig2} (c) and (d) respectively where we used the coupling $\tilde{g}/2\pi = 35$~MHz for illustration purposes, which is within the same order as the coupling reported in Ref.~\cite{Harder2018}.

Experimentally, a coupling between a microwave cavity field with externally driven magnetization has been realized in Refs.~\cite{Boventer2019, Boventer2020}. These experiments demonstrated level attraction at substantially large driving field strengths.  We think these results may be interpreted in terms of coupling of microwave cavity photons with excitations above a nonlinear stationary state established by driving.  In this case, the non-Hermitian coupling introduced in Refs.~\cite{Boventer2019, Boventer2020} corresponds in our picture to the coupling to the lower side-band resonance in Eq.~(\ref{ps2}). However, a detailed discussion requires additional analysis since the driving field used in Refs.~\cite{Boventer2019, Boventer2020} is linearly polarized.

%% ============================================================================
\section{Conclusion}
\label{SecV}
We consider the possibility of  level attraction in a coupled cavity magnon-polariton system, where a magnetic system is driven independently out of equilibrium.  Using a master equation formalism, we demonstrate that this problem can be analyzed from the broader perspective of Floquet dynamics in systems with memory \cite{Traversa2013, Magazzu2017}. From this point of view, level attraction can be interpreted as an instability developed as a result of interaction between excitations above a stationary driven magnetization \textbf{P}-mode and a probe system of cavity photons.  We show that this instability develops when the cavity resonance is close to the frequency of the lower side band around the \textbf{P}-mode. The resulting interaction between two resonances can be interpreted using an effective non-Hermitian Hamiltonian with a dissipative coupling term \cite{Grigoryan2019, Boventer2020}. Level attraction quickly disappears when the reservoir approaches thermodynamic equilibrium. Our approach is promising for future analysis of level attraction in Floquet cavity magnonics \cite{Xu2020} with non-equilibrium excited steady states such as discrete magnetic breather modes, and will be explained in future work.
%% ============================================================================

\begin{acknowledgments}
RLS acknowledges the support from the Natural Sciences and Engineering Research Council of Canada (NSERC) RGPIN 05011-18, the Canada Foundation for Innovation JELF and the University of Manitoba.
\end{acknowledgments}

\onecolumngrid
\appendix

\section{Master equation for open systems}
\label{app_A}

By applying $\Sb$ to both sides of Eq.~(\ref{liouv}), we obtain (note that the source term disappears under this transformation)
\begin{equation}  \label{proj}
	\left( \frac{\partial}{\partial t} + i\mathcal{L}_{s}^{t}  \right) \hat{\rho}(t) = -\Sb \left( i\mathcal{L}^{t}_{i} \hat{\rho}_{\mathrm{tot}}(t)  \right).
\end{equation}
To close this equation, we introduce $\Delta \hat{\rho}(t)$ using the following definition
\begin{equation}  \label{drho}
	\hat{\rho}_{\mathrm{tot}}(t) = \hat{\rho}(t) \hat{\rho}_{r}(t) + \Delta \hat{\rho}(t),
\end{equation}
with the boundary condition $\Delta \hat{\rho}(t) = 0$ at $t \to -\infty$. From Eqs.~(\ref{liouv}) and (\ref{proj}), after a little algebra, we find that 
\begin{equation}\label{lvl1}
	\left( \frac{\partial}{\partial t} + i\mathcal{L}_{s}^{t} + i\mathcal{L}_{r}^{t} + i\mathcal{L}^{t}_{i} + \varepsilon  \right) \Delta \hat{\rho}(t) = -i\mathcal{L}^{t}_{i} \hat{\rho}(t) \hat{\rho}_{r}(t) + \Sb\left[ i\mathcal{L}^{t}_{i} \hat{\rho}_{\mathrm{tot}}(t) \right] \hat{\rho}_{r}(t) 
	- \hat{\rho}(t) \left( \frac{\partial \hat{\rho}_{r}(t)}{\partial t} + i\mathcal{L}_{r}^{t} \hat{\rho}_{r}(t)  \right).
\end{equation}
The last term on the right hand side vanishes, since $\hat{\rho}_{r}(t)$ satisfies Eq.~(\ref{lvl}).  This equation contains $\Delta \hat{\rho}(t)$ in the second term at the right hand side.

To deal with the second term, we introduce a projection operator
\begin{equation}
	\Pj \hat{A} = \hat{\rho}_{r}(t) \Sb \hat{A},
\end{equation}
where $\hat{A}$ is any operator in the Hilbert space of the whole system. Note that since we consider the reservoir to be dynamics, $\Pj$ can bring additional time dependence. With the help of this definition, we find
\begin{equation}
	\Sb\left( i\mathcal{L}^{t}_{i} \hat{\rho}_{\mathrm{tot}}(t)  \right) \hat{\rho}_{r}(t) = \Pj i\mathcal{L}^{t}_{i} \hat{\rho}_{\mathrm{tot}}(t) = \Pj i\mathcal{L}^{t}_{i} \hat{\rho}(t)  \hat{\rho}_{r}(t) + \Pj i\mathcal{L}^{t}_{i} \Delta \hat{\rho}(t). 
\end{equation}
This allows to rewrite Eq.~(\ref{lvl1}) in the following form
\begin{equation}
	\left( \frac{\partial}{\partial t} + i\mathcal{L}_{s}^{t} + i\mathcal{L}_{r}^{t} + \Qj i\mathcal{L}^{t}_{i} \Qj + \varepsilon  \right) \Delta \hat{\rho}(t) = -i\mathcal{L}^{t}_{i} \hat{\rho}(t) \hat{\rho}_{r}(t) + \Pj i\mathcal{L}^{t}_{i} \hat{\rho}(t) \hat{\rho}_{r}(t),
\end{equation}
where $\Qj = 1 - \Pj$.

Finally, the right hand side of this equation can be simplified using the following identity
\begin{equation}
	\Sb i\mathcal{L}^{t}_{i} \hat{\rho}(t) \hat{\rho}_{r}(t) = \frac{1}{i\hbar} \left[ \hat{\rho}(t), \Sb\left( \hat{H}^{t}_{i} \hat{\rho}_{r}(t)  \right)  \right],
\end{equation}
which gives
\begin{equation} \label{drho2}
	\left( \frac{\partial}{\partial t} + i\mathcal{L}_{s}^{t} + i\mathcal{L}_{r}^{t} + \Qj i\mathcal{L}^{t}_{i} \Qj + \varepsilon  \right) \Delta \hat{\rho}(t) = -i\Delta \mathcal{L}^{t}_{i} \hat{\rho}(t) \hat{\rho}_{r}(t). 
\end{equation}
where the Liouville operator at the right hand side is defined as follows
\begin{equation}
	i\Delta \mathcal{L}^{t}_{i} \ldots = \frac{1}{i\hbar} \left[\ldots, \Delta\hat{H}^{t}_{i} \right], 
	\quad \mbox{where} \quad
	\Delta \hat{H}^{t}_{i} = \hat{H}^{t}_{i} - \Sb\left( \hat{H}^{t}_{i} \hat{\rho}_{r}(t) \right).
\end{equation}

\subsection{Formal solution for $\Delta \hat{\rho}(t)$ and master equation for $\hat{\rho}(t)$}

Formal solution of the Liouville equation (\ref{drho2}) is 
\begin{equation}
	\Delta \hat{\rho}(t) = -\int_{-\infty}^{t} dt' \mathcal{U}_{tt'} i\Delta \mathcal{L}^{t'}_{i} \hat{\rho}(t') \hat{\rho}_{r}(t'),
\end{equation}
and the evolution operator is given by
\begin{equation}
	\mathcal{U}_{tt'} = \mathcal{T} \exp\left[ -\int_{t'}^{t} d\tau \left( i\mathcal{L}_{s}^{\tau} + i\mathcal{L}_{r}^{\tau} + \mathcal{Q}_{\tau} i\mathcal{L}^{\tau}_{i} \mathcal{Q}_{\tau} + \varepsilon \right) \right], \qquad t > t',
\end{equation} 
where $\mathcal{T}$ is the time ordering operator.

This allows us to write down an equation for $\hat{\rho}(t)$ in a closed form
\begin{multline} \label{rho100}
	\left( \frac{\partial}{\partial t} + i\mathcal{L}_{s}^{t}  \right) \hat{\rho}(t) + \Sb \left[ i\mathcal{L}^{t}_{i} \hat{\rho}(t) \hat{\rho}_{r}(t) \right]  \\ 
	= \int_{-\infty}^{t} dt' e^{-\varepsilon(t-t')}\Sb i\mathcal{L}^{t}_{i} \mathcal{T} \exp\left\lbrace -\int_{t'}^{t} d\tau \left( i\mathcal{L}_{s}^{\tau} + i\mathcal{L}_{r}^{\tau} + \mathcal{Q}_{\tau} i\mathcal{L}^{\tau}_{i} \mathcal{Q}_{\tau} \right)  \right\rbrace  i\Delta \mathcal{L}^{t'}_{i} \hat{\rho}(t') \hat{\rho}_{r}(t').
\end{multline}
The fact that $\Sb \left(\Delta \hat{\rho} (t) \right) = 0$ allows us to replace $i\mathcal{L}^{t}_{i}$ with $i\Delta \mathcal{L}^{t}_{i}$ in this equation, so that it would have a symmetric form.

In the weak interaction limit, a great simplification is achieved by neglecting the higher order interaction terms, $\mathcal{Q}_{\tau} i\mathcal{L}^{\tau}_{i} \mathcal{Q}_{\tau}$, inside the argument of the exponent at the right hand side of Eq.~(\ref{rho100}), which also disentangles evolution operators for the probe system and reservoir. In this approximation, the right hand side is of the second order in $\hat{H}_{i}^{t}$.

Now, by transforming the Liouville operators into the commutators, we can rewrite the master equation for $\hat{\rho}(t)$ the following form
\begin{equation}
	\frac{\partial \hat{\rho}(t)}{\partial t} + \frac{1}{i\hbar} \left[\hat{\rho}(t), \hat{H}_{s}^{t} + \Sb\left( \hat{H}^{t}_{i} \hat{\rho}_{r}(t)  \right) \right]
	= \frac{1}{(i\hbar)^{2}} \int_{-\infty}^{t} dt' e^{-\varepsilon (t - t')} \Sb \left[ \hat{H}^{t}_{i}, \hat{U}_{tt'} \left[ \Delta \hat{H}^{t'}_{i}, \hat{\rho}(t') \hat{\rho}_{r}(t') \right] \hat{U}_{tt'}^{-1}  \right],
\end{equation}
where the evolution operator is defined in Eq.~(\ref{Uev}).

The next simplification is reached by observing that dynamics of $\hat{\rho}_{r}(t)$ is independent from dynamics of the probe, and satisfies equation (\ref{lvl}), so that $\hat{U}_{tt'} \hat{\rho}_{r}(t') \hat{U}_{tt'}^{-1} =  \hat{\rho}_{r}(t)$.

By introducing shorthand notations $\hat{\rho}(t,t') = \hat{U}_{tt'} \hat{\rho}(t') \hat{U}_{tt'}^{-1}$ and $\Delta \hat{H}^{t'}_{i}(t,t') = \hat{U}_{tt'} \Delta \hat{H}^{t'}_{i} \hat{U}_{tt'}^{-1}$, we rewrite the master equation for $\hat{\rho}(t)$ in the weak interaction limit in the form of Eq.~(\ref{mas}).

\subsection{Kinetic equations for dynamic variables}
Further simplification is reached by considering the interaction Hamiltonian in the following form $\hat{H}_{i} = \hat{A} \hat{F} + \hat{A}^{\dag} \hat{F}^{\dag}$, where the operators $\hat{A}$ and $\hat{A}^{\dag}$ act in the Hilbert space of the probe system and $\hat{F}$ and $\hat{F}^{\dag}$ act entirely on the degrees of freedom of the reservoir. In this case, $ \Delta \hat{H}^{t'}_{i} = \hat{A} \Delta \hat{F}_{t'} + \hat{A}^{\dag} \Delta \hat{F}^{\dag}_{t'} $, where $\Delta \hat{F}_{t'} = \hat{F} - \langle \hat{F} \rangle^{t'}_{r}$, and $\langle\hat{F} \rangle^{t'}_{r} = \Sb\left[ \hat{F} \hat{\rho}_{r}(t') \right]$.  We do not specify any specific commutation rules for $\hat{A}$ and $\hat{A}^{\dag}$ at this stage.

The explicit form of the master equation for $\hat{\rho}(t)$ is obtained straightforwardly from Eq.~(\ref{mas})
\begin{eqnarray}
	\nonumber
	\frac{\partial \hat{\rho}(t)}{\partial t} + \frac{1}{i\hbar} \left[ \hat{\rho}(t), \hat{H}_{s}^{t}  \right] + \frac{1}{i\hbar} \left[ \hat{\rho}(t), \hat{A}  \right] \langle \hat{F} \rangle_{r}^{t} + \frac{1}{i\hbar} \left[ \hat{\rho}(t), \hat{A}^{\dag}  \right] \langle \hat{F}^{\dag} \rangle_{r}^{t} 
	= \frac{1}{(i\hbar)^{2}} \int_{-\infty}^{t} dt' e^{-\varepsilon(t - t')} \\
	\nonumber
	\times
	\left\lbrace
	\langle \hat{F} \Delta \hat{F}_{t'}(t,t')  \rangle_{r}^{t} \hat{A} \hat{A}(t,t') \hat{\rho}(t,t') + 
	\langle \hat{F}^{\dag} \Delta \hat{F}^{\dag}_{t'}(t,t')  \rangle_{r}^{t} \hat{A}^{\dag} \hat{A}^{\dag}(t,t') \hat{\rho}(t,t') 
	+ \langle \hat{F} \Delta \hat{F}^{\dag}_{t'}(t,t')  \rangle_{r}^{t} \hat{A} \hat{A}^{\dag}(t,t') \hat{\rho}(t,t') 
	\right.
	\\
	\nonumber
	\left.
	+ 
	\langle \hat{F}^{\dag} \Delta \hat{F}_{t'}(t,t')  \rangle_{r}^{t} \hat{A}^{\dag} \hat{A}(t,t') \hat{\rho}(t,t') 
	+ \langle \Delta \hat{F}_{t'}(t,t') \hat{F} \rangle_{r}^{t} \hat{\rho}(t,t') \hat{A}(t,t') \hat{A} 
	+ \langle \Delta \hat{F}^{\dag}_{t'}(t,t') \hat{F}^{\dag} \rangle_{r}^{t} \hat{\rho}(t,t') \hat{A}^{\dag}(t,t') \hat{A}^{\dag} \right.
	\\
	\nonumber
	\left.
	+ \langle \Delta \hat{F}_{t'}(t,t') \hat{F}^{\dag} \rangle_{r}^{t} \hat{\rho}(t,t') \hat{A}(t,t') \hat{A}^{\dag} + \langle \Delta \hat{F}^{\dag}_{t'}(t,t') \hat{F} \rangle_{r}^{t} \hat{\rho}(t,t') \hat{A}^{\dag}(t,t') \hat{A} 
	-\langle \Delta \hat{F}_{t'}(t,t') \hat{F} \rangle_{r}^{t} \hat{A} \hat{\rho}(t,t')  \hat{A}(t,t') 
	\right.
	\\
	\nonumber
	\left.
	- 
	\langle \Delta \hat{F}^{\dag}_{t'}(t,t') \hat{F}^{\dag} \rangle_{r}^{t} \hat{A}^{\dag}  \hat{\rho}(t,t') \hat{A}^{\dag}(t,t') 
	- \langle \Delta \hat{F}^{\dag}_{t'}(t,t') \hat{F} \rangle_{r}^{t} \hat{A} \hat{\rho}(t,t') \hat{A}^{\dag}(t,t') - 
	\langle \Delta \hat{F}_{t'}(t,t') \hat{F}^{\dag} \rangle_{r}^{t} \hat{A}^{\dag} \hat{\rho}(t,t') \hat{A}(t,t') \right.
	\\
	\nonumber
	\left.
	- \langle  \hat{F} \Delta \hat{F}_{t'}(t,t') \rangle_{r}^{t} \hat{A}(t,t') \hat{\rho}(t,t') \hat{A} - \langle \hat{F}^{\dag} \Delta \hat{F}^{\dag}_{t'}(t,t') \rangle_{r}^{t} \hat{A}^{\dag}(t,t') \hat{\rho}(t,t') \hat{A}^{\dag} 
	- \langle \hat{F}^{\dag} \Delta \hat{F}_{t'}(t,t') \rangle_{r}^{t} \hat{A}(t,t') \hat{\rho}(t,t') \hat{A}^{\dag} 
	\right.
	\\
	\left.
	- \langle \hat{F} \Delta \hat{F}^{\dag}_{t'}(t,t') \rangle_{r}^{t} \hat{A}^{\dag}(t,t') \hat{\rho}(t,t')  \hat{A} 
	\right\rbrace,
	\label{rho300}
\end{eqnarray}
where we used shorthand notations $ \hat{A}(t,t')  = \hat{U}_{tt'} \hat{A} \hat{U}_{tt'}^{-1} $ and $ \hat{F}(t,t')  = \hat{U}_{tt'} \hat{F} \hat{U}_{tt'}^{-1} $, and $\langle \ldots \rangle_{r}^{t} \equiv \Sb[\ldots \hat{\rho}_{r}(t)]$.

Correlation functions between $\hat{F}$ and $\hat{F}^{\dag}$ in this equations can be transformed to a more physically transparent form. For this purpose, we introduce the Heisenberg picture for operators as follows
\begin{equation} \label{hein}
	\hat{A}(t) = \hat{U}^{-1}_{t,-\infty} \hat{A} \hat{U}_{t,-\infty},
\end{equation}
where the explicit expressions for the evolution operators are given by 
\begin{equation} \label{evl01}
	\hat{U}_{t,-\infty} = \mathcal{T}\exp\left( -\frac{i}{\hbar} \int_{-\infty}^{t} \hat{H}_{r}^{\tau} d\tau \right) \quad \mbox{and} \quad
	\hat{U}_{t,-\infty}^{-1} = \mathcal{T}_{a}\exp\left( \frac{i}{\hbar} \int_{-\infty}^{t} \hat{H}_{r}^{\tau} d\tau  \right).
\end{equation}

In this notation, we can express these correlation functions in terms of nonequilibrium Green functions \cite{Zubarev1996}
\begin{equation} \label{id100}
	\langle \hat{F}^{\dag} \hat{F}(t,t') \rangle_{r}^{t} = \langle \hat{F}^{\dag}(t) \hat{F}(t') \rangle_{r} \equiv \Sb \left\lbrace \hat{F}^{\dag}(t) \hat{F}(t')\rho_{r}(-\infty) \right\rbrace, \quad t > t'.
\end{equation}

We now apply the master equation for $\hat{\rho}(t)$ to derive kinetic equations for dynamic variables. For a dynamics variable described by a general operator $\hat{B}$ in the Hilbert space of the probe system, we define the average value at time moment $t$ as $\langle \hat{B} \rangle^{t} = \Tr \left[ \hat{B} \hat{\rho}(t)  \right]$. In this case using Eq.~(\ref{rho300}), we find
\begin{eqnarray}
	\nonumber
	\frac{d}{dt} \langle \hat{B} \rangle^{t} + \frac{1}{i\hbar} \left\langle\left[        \hat{H}_{s}^{t}, \hat{B}
	\right]\right\rangle^{t} + \frac{1}{i\hbar} \langle \hat{F}^{\dag}  \rangle^{t}_{r}  \left\langle \left[
	\hat{A}^{\dag}, \hat{B}
	\right]\right\rangle^{t} + \frac{1}{i\hbar} \langle \hat{F}  \rangle^{t}_{r}  \left\langle \left[
	\hat{A}, \hat{B}
	\right]\right\rangle^{t} = \frac{1}{(i\hbar)^{2}} \int_{-\infty}^{t} dt' e^{-\varepsilon(t - t')} \\
	\nonumber
	\times
	\left\lbrace
	\langle \hat{F} \Delta \hat{F}_{t'}(t,t')  \rangle_{r}^{t} 
	\langle [\hat{B}(t',t), \hat{A}(t',t)] \hat{A} \rangle^{t'} + 
	\langle \hat{F}^{\dag} \Delta \hat{F}^{\dag}_{t'}(t,t')  \rangle_{r}^{t} 
	\langle [\hat{B}(t',t), \hat{A}^{\dag}(t',t)] \hat{A}^{\dag} \rangle^{t'}  \right.
	\\
	\nonumber
	\left.
	+ \langle \hat{F} \Delta \hat{F}^{\dag}_{t'}(t,t')  \rangle_{r}^{t} 
	\langle [\hat{B}(t',t), \hat{A}(t',t)] \hat{A}^{\dag} \rangle^{t'}  + 
	\langle \hat{F}^{\dag} \Delta \hat{F}_{t'}(t,t')  \rangle_{r}^{t} 
	\langle [\hat{B}(t',t), \hat{A}^{\dag}(t',t)] \hat{A} \rangle^{t'}  \right.
	\\
	\nonumber
	\left.
	+ \langle \Delta \hat{F}_{t'}(t,t') \hat{F} \rangle_{r}^{t} 
	\langle \hat{A} [\hat{A}(t',t), \hat{B}(t',t)] \rangle^{t'} 
	+ \langle \Delta \hat{F}^{\dag}_{t'}(t,t') \hat{F}^{\dag} \rangle_{r}^{t} 
	\langle \hat{A}^{\dag} [\hat{A}^{\dag}(t',t), \hat{B}(t',t)] \rangle^{t'} 
	\right.
	\\ \label{gkin}
	\left.
	+ \langle \Delta \hat{F}_{t'}(t,t') \hat{F}^{\dag} \rangle_{r}^{t} 
	\langle \hat{A} [\hat{A}^{\dag}(t',t), \hat{B}(t',t)] \rangle^{t'} 
	+ \langle \Delta \hat{F}^{\dag}_{t'}(t,t') \hat{F} \rangle_{r}^{t} 
	\langle \hat{A}^{\dag} [\hat{A}(t',t), \hat{B}(t',t)] \rangle^{t'}
	\right\rbrace,
\end{eqnarray}
where $\hat{B}(t',t) = \hat{U}_{tt'}^{-1} \hat{B} \hat{U}_{tt'}$.

In what follows, we will be interested in a situation when $\hat{B}$ is the same dynamic variable as in $\hat{H}_{i}$, i.~e. $\hat{B} = \hat{A}$.  In this case, the kinetic equation is simplified
\begin{eqnarray}
	\nonumber
	\frac{d}{dt} \langle \hat{A} \rangle^{t} + \frac{1}{i\hbar} \left\langle\left[        \hat{H}_{s}^{t}, \hat{A}
	\right]\right\rangle^{t} + 
	\frac{1}{i\hbar} \langle \hat{F}^{\dag}  \rangle^{t}_{r}  \left\langle \left[
	\hat{A}^{\dag}, \hat{A}
	\right]\right\rangle^{t} 
	= \frac{1}{(i\hbar)^{2}} \int_{-\infty}^{t} dt' e^{-\varepsilon(t - t')} \\
	\nonumber
	\times
	\left\lbrace
	\langle \hat{F}^{\dag} \Delta \hat{F}^{\dag}_{t'}(t,t')  \rangle_{r}^{t} 
	\langle [\hat{A}(t',t), \hat{A}^{\dag}(t',t)] \hat{A}^{\dag} \rangle^{t'} 
	+\langle \hat{F}^{\dag} \Delta \hat{F}_{t'}(t,t')  \rangle_{r}^{t} 
	\langle [\hat{A}(t',t), \hat{A}^{\dag}(t',t)] \hat{A} \rangle^{t'}  \right.
	\\
	\left.
	+\langle \Delta \hat{F}^{\dag}_{t'}(t,t') \hat{F}^{\dag} \rangle_{r}^{t} 
	\langle \hat{A}^{\dag} [\hat{A}^{\dag}(t',t), \hat{A}(t',t)] \rangle^{t'} 
	+\langle \Delta \hat{F}_{t'}(t,t') \hat{F}^{\dag} \rangle_{r}^{t} 
	\langle \hat{A} [\hat{A}^{\dag}(t',t), \hat{A}(t',t)] \rangle^{t'} 
	\right\rbrace.
\end{eqnarray}

And finally, if $\hat{A} = \hat{a}$, where $\hat{a}$ and $\hat{a}^{\dag}$ satisfy the boson commutation rules, $[\hat{a}, \hat{a}^{\dag}] = 1$, and we obtain
\begin{equation}
	\nonumber
	\frac{d \langle \hat{a} \rangle^{t} }{dt} + \frac{1}{i\hbar} \left\langle\left[        \hat{H}_{s}^{t}, \hat{a}
	\right]\right\rangle^{t} - 
	\frac{1}{i\hbar} \langle \hat{F}^{\dag}  \rangle^{t}_{r}
	= \frac{1}{(i\hbar)^{2}} \int_{-\infty}^{t} dt' e^{-\varepsilon(t - t')} 
	\left\lbrace
	\langle [\hat{F}^{\dag},  \hat{F}^{\dag}(t,t')]  \rangle_{r}^{t} 
	\langle \hat{a}^{\dag} \rangle^{t'} 
	+\langle [\hat{F}^{\dag}, \hat{F}(t, t')] \rangle_{r}^{t} 
	\langle \hat{a} \rangle^{t'} 
	\right\rbrace.
\end{equation}
A contribution from the off-resonant term, proportional to $ \langle [\hat{F}^{\dag}(t),  \hat{F}^{\dag}(t')]  \rangle_{r}$,  is usually small compared to $ \langle [\hat{F}^{\dag}(t),  \hat{F}(t')]  \rangle_{r}$, which describes resonant interaction between the probe system and reservoir. If $ \omega_{0} $ is a characteristic frequency of the reservoir, the off-resonant term oscillates at $ 2\omega_{0} $, and can be neglected when $ \omega \tau_{r} \gg 1$, where $ \tau_{r} $ is a characteristic relaxation time of the reservoir \cite{Zubarev1996}.

If we neglect the off-resonant term proportional to $\langle \hat{a}^{\dag} \rangle^{t}$ and use the identity in Eq.~(\ref{id100}), we obtain the following kinetic equation for $\langle \hat{a} \rangle^{t}$
\begin{equation} %\label{kin2}
	i\hbar\frac{d \langle \hat{a} \rangle^{t} }{dt} 
	+ \left\langle\left[\hat{H}_{s}^{t}, \hat{a} \right]\right\rangle^{t} 
	- \int_{-\infty}^{t} dt'
	G(t,t') \langle \hat{a} \rangle^{t'} =
	\langle \hat{F}^{\dag}  \rangle^{t}_{r},
\end{equation}
where we introduce nonequilibrium retarded Green function
\begin{equation} \label{kin1}
	G(t,t') = \frac{e^{-\varepsilon(t-t')}}{i\hbar}\theta(t-t') \langle [\hat{F}^{\dag}(t), \hat{F}(t')] \rangle_{r},
\end{equation}
where the operators $ \hat{F} $ and $ \hat{F}^{\dag} $ are in the Heisenberg picture as defined by Eq~(\ref{hein}), and $\langle \ldots \rangle_{r} \equiv \Sb\left(\ldots \hat{\rho}_{r}(-\infty)\right)$.

\section{Perturbative expansion over a \textbf{P}-mode solution}
\label{app_B}

Here, we discuss how to calculate spin-spin Poisson brackets for linear excitations around a stationary \textbf{P}-mode trajectory. The Poisson bracket for two circularly polarized spin components $ S^{(\pm)} $ taken at different time moments is defined in Eq.~(\ref{psn}). For this purpose, we expand the equations of motion (\ref{eom10}) and (\ref{eom11}) over the stationary
\textbf{P}-mode solution: $\theta(t) = \theta_{0} + \delta \theta(t)$ and 
$\phi(t) = \phi_{0} + \delta \phi(t)$, where $\theta_{0}$ and $\phi_{0}$
denote stationary solution defined in Eq.~(\ref{st10}) and (\ref{st11}).  To linear order in $\delta \phi(t)$ and $\delta \theta(t)$, the equations of motion become
\begin{eqnarray} \label{B1}
	\delta\dot{\theta} - \alpha\sin\theta_{0} \delta\dot{\phi} &=&
	\varkappa \left(b_{\perp}\cos\phi_{0}\delta\phi - \Omega\cos\theta_{0} \delta\theta \right), \\ \label{B2}
	\alpha\delta\dot{\theta} + \sin\theta_{0} \delta\dot{\phi} &=& 
	-\varkappa \left( b_{\perp} \sin\phi_{0}\cos\theta_{0} \delta\phi  
	+ b_{\perp} \cos\phi_{0}\sin\theta_{0} \delta\theta 
	+ b_{z} \cos\theta_{0} \delta\theta + \cos2\theta_{0} \delta\theta \right).
\end{eqnarray}
These equations are also given in the matrix form in Eq.~(\ref{eom}).
Two important characteristics of this equation are the trace and the determinant of the matrix on the right hand side of Eq.~(\ref{eom}) (denoted here as $M$):
\begin{equation} \label{a100}
	\Tr M = -\frac{2\alpha \varkappa}{1 + \alpha^{2}}
	\left(
	v -\frac{1-m_{z}^{2}}{2} + \frac{\Omega m_{z}}{\alpha}
	\right), \quad
	\det M = \frac{\varkappa^{2}} {1 + \alpha^{2}}
	\left(v^{2} - (1 - m_{z}^{2})v + \Omega^{2}m_{z}^{2}\right),
\end{equation}
from which a phase diagram for \textbf{P}-mode stability can be obtained \cite{Bertotti2001}.

 Let us first illustrate how to calculate this Poisson bracket in absence of anisotropy, $ \varkappa=0 $, and dissipation, $ \alpha=0 $. In this case, the  equations of motion in (\ref{B1}) and (\ref{B2}) reduce for those of a harmonic oscillator
\begin{equation} \label{zal}
	\delta \dot{\theta} = h_{a\perp} \delta \phi, \qquad
	\delta \dot{\phi} = -\frac{h_{a\perp}}{\sin^{2}\theta_{0}} \delta \theta.
\end{equation}
Solutions for these equation, which satisfy the initial conditions $\delta \phi(0) = \delta \phi_{0}$ and $\delta \theta(0) = \delta \theta_{0}$, have the following form
\begin{eqnarray} \label{sol10}
	\delta \theta(t) &=& \delta \theta_{0} \cos\omega_{0}t + \frac{h_{a\perp}}{\tilde{\omega}_{0}} \delta \phi_{0} \sin\omega_{0}t,\\ \label{sol11}
	\delta \phi(t) &=& -\frac{\tilde{\omega}_{0}}{h_{a\perp}} \delta\theta_{0} \sin \omega_{0}t + \delta \phi_{0} \cos\omega_{0} t,
\end{eqnarray}
where $\tilde{\omega}_{0} = h_{a\perp}/\sin\theta_{0} $ denotes the dimensionless frequency.

In this situation, $ S^{(+)}(t) $ is expanded around the \textbf{P}-mode as follows
\begin{equation} \label{po100}
	S^{(+)} (t) = \sin \theta_{0} e^{i\omega' t} + e^{i(\omega' + \omega_{0})t} 
	\cos^{2} \frac{\theta_{0}}{2} \left( \delta\theta_{0} - i\sin\theta_{0} \delta \phi_{0}  \right) 
	- e^{i(\omega' - \omega_{0})t} \sin^{2} \frac{\theta_{0}}{2}  \left( \delta \theta_{0} + i\sin \theta_{0} \delta \phi_{0} \right).
\end{equation}
From the definition of the Poisson bracket in Eq.~(\ref{psn}),  we restore Eq.~(\ref{ps2}) in the main text. Note that in the limit of $\theta_{0} \to 0$, this expression reduces to $\lbrace S^{(+)}(t), S^{(-)}(0) \rbrace \to -2i e^{i(\omega' + \omega_{0})t}$,  where $\omega_{0} \approx \gamma B_{z} - \omega'$, which corresponds to the usual ferromagnetic resonance with the frequency $\gamma B_{z}$.

The explicit expression for the spin-spin Poisson bracket in the general case is given by
\begin{equation} \label{eom100}
	\lbrace S^{(+)}(t), S^{(-)}(0) \rbrace = \frac{e^{i\omega' t}}{\sin\theta_{0}}
	\left[ \cos^{2}\theta_{0} \frac{\partial \delta \theta(t)}{\partial \delta \phi_{0}}  
	-\sin^{2}\theta_{0} \frac{\partial \delta \phi(t)}{\partial \delta \theta_{0}} 
	-i\sin\theta_{0} \cos \theta_{0}\left( \frac{\partial \delta \phi(t)}{\partial \delta \phi_{0}} + \frac{\partial \delta \theta(t)}{\partial \delta \theta_{0}}  \right) \right],
\end{equation}
where $\delta\phi(t)$ and $\delta\theta(t)$ satisfy Eq.~(\ref{eom}) with $ \delta \theta(0) = \delta \theta_{0} $ and $ \delta \phi(0) = \delta\phi_{0} $. A solution of Eq.~(\ref{eom}) that satisfies these initial conditions can be written in the following general form
\begin{equation} 
	\begin{pmatrix}
		\delta \theta(t) \\
		\delta \phi(t)
	\end{pmatrix}
	= e^{-\gamma t} \left[
	\frac{\delta \theta_{0} v_{2} - \delta\phi_{0}v_{1}}{u_{1}v_{2} - u_{2}v_{1}}
	\begin{pmatrix}
		u_{1} \\ u_{2}
	\end{pmatrix}
	e^{i\omega_{0}t}
	+
	\frac{\delta \phi_{0} u_{1} - \delta\theta_{0} u_{2}}{u_{1}v_{2} - u_{2}v_{1}}
	\begin{pmatrix}
		v_{1} \\ v_{2}
	\end{pmatrix}
	e^{-i\omega_{0}t}
	\right],
\end{equation}
where $(u_{1},u_{2})^{T}$ is the eigenvector (not necessary normalized) that corresponds to $\lambda_{+} = -\gamma + i\omega_0$ and $(v_{1},v_{2})^{T}$ corresponds to $\lambda_{-} = -\gamma - i\omega_0$. The explicit expressions (in dimensionless units) can be found from Eq.~(\ref{eom}) and (\ref{a100})
\begin{equation}
	\tilde{\lambda}_{\pm} = \frac{1}{2}\left( \tr M \pm \sqrt{(\tr M)^{2} - 4\det M}\right) \equiv -\tilde{\gamma} \pm i\tilde{\omega}_{0},
\end{equation}
where $\tilde{\gamma} = -\frac{1}{2} \tr M$, and $i\tilde{\omega}_{0} = \sqrt{(\tr M)^{2} - 4\det M}$.
For components of the eigenvectors we have
\begin{equation} \label{uv}
	\begin{pmatrix}
		u_{1} & v_{1} \\
		u_{2} & v_{2}
	\end{pmatrix}
	= 
	\begin{pmatrix}
		- \dfrac{\sin \theta_{0}(v-\alpha\Omega m_{z})}{\frac{1}{2}\alpha(1-m_{z}^{2})-i\tilde{\omega_{0}}} &
		-\dfrac{\sin \theta_{0}(v-\alpha\Omega m_{z})}{\frac{1}{2}\alpha(1-m_{z}^{2})+i\tilde{\omega_{0}}} \\
		1 & 1
	\end{pmatrix}.
\end{equation}

By expanding $ S^{(+)}(t) $ around the stationary solution and calculating the derivatives with respect to the initial condition in Eq.~(\ref{eom100}), we obtain
\begin{equation} \label{com1}
	\lbrace S^{(+)}(t), S^{(-)}(0) \rbrace =
	-2i \frac{e^{i\omega' t - \gamma t}}{\sin\theta_{0}} \left[
	\left( \frac{u_{1}v_{1}}{\Delta} \cos^{2}\theta_{0} + \frac{u_{2}v_{2}}{\Delta} \sin^{2}\theta_{0} \right)\sin\omega_{0}t + \sin \theta_{0}\cos\theta_{0} \cos\omega_{0}t
	\right],
\end{equation}
where $\Delta = u_{1}v_{2} - u_{2}v_{1}$.

From equations (\ref{uv})--(\ref{com1}), we finally obtain the Poisson bracket in the following form
\begin{multline} \label{po300}
	\lbrace S^{(+)}(t), S^{(-)}(0) \rbrace =-\frac{ie^{-\gamma t}}{4\tilde{\omega}_{0}(v-\alpha\Omega m_z)}\left\lbrace
	\left[
	\left((v-\alpha\Omega m_{z})m_{z} + \tilde{\omega}_{0}\right)^{2} + \frac{1}{4}\alpha^{2}(1-m_{z}^{2})^{2}
	\right]e^{i(\omega' + \omega_{0})t} 
	\right. \\
	- 
	\left.\left[
	\left((v-\alpha\Omega m_{z})m_{z} - \tilde{\omega}_{0}\right)^{2} + \frac{1}{4}\alpha^{2}(1-m_{z}^{2})^{2}
	\right]e^{i(\omega' - \omega_{0})t} 
	\right\rbrace.
\end{multline}
Note that each term in the square brackets is manifestly positive.  When $\varkappa \to 0$, we have $\tilde{\omega}_{0} \to  v-\alpha\Omega m_{z}$, and this equation reduces to Eq.~(\ref{ps2}) in the main text of the paper.

\twocolumngrid
\bibliography{noneql}

\end{document}